\documentclass[%
 aip,
 amsmath,amssymb,
 reprint,%
]{revtex4-1}

\usepackage{graphicx}
\usepackage{dcolumn}
\usepackage{bm}

\usepackage{amssymb}
\usepackage{amsmath}
\usepackage{times}
\usepackage[outdir=./]{epstopdf}
\usepackage{lipsum, babel}
\usepackage{xcolor} 
\usepackage{ulem}

\definecolor{internationalkleinblue}{rgb}{0.0, 0.18, 0.65}
\definecolor{brickred}{rgb}{0.8, 0.25, 0.33}
\definecolor{darkviolet}{rgb}{0.58, 0.0, 0.83}

\usepackage[utf8]{inputenc}
\usepackage[T1]{fontenc}
\usepackage{mathptmx}
\usepackage{etoolbox}

\makeatletter
\def\@email#1#2{%
 \endgroup
 \patchcmd{\titleblock@produce}
  {\frontmatter@RRAPformat}
  {\frontmatter@RRAPformat{\produce@RRAP{*#1\href{mailto:#2}{#2}}}\frontmatter@RRAPformat}
  {}{}
}%
\makeatother
\begin{document}

\preprint{AIP/123-QED}

\title[Higher-order interactions induce anomalous transitions to synchrony]{Higher-order interactions induce anomalous transitions to synchrony}

\author{Iv\'an Le\'on}
\affiliation{Department of Systems and Control Engineering, Tokyo Institute of Technology, Tokyo 152-8550, Japan}
\email{leonmerinoi@unican.es}
\author{Riccardo Muolo}
\affiliation{Department of Mathematics and naXys, Namur Institute for Complex Systems, University of Namur, Rue Grafé 2, 5000 Namur, Belgium}
\author{Shigefumi Hata}
\affiliation{Graduate School of Science and Engineering, Kagoshima University, Korimoto 1-21-35, 890-0065 Kagoshima, Japan}
\author{Hiroya Nakao}
\affiliation{Department of Systems and Control Engineering, Tokyo Institute of Technology, Tokyo 152-8550, Japan}

\date{\today}

\begin{abstract}

We analyze the simplest model of identical coupled phase oscillators subject to two-body and three-body interactions with permutation symmetry {  and phase lags}. This model is derived from an ensemble of weakly coupled nonlinear oscillators by phase reduction{ ,  where 
the first and second harmonic interactions with phase lags naturally appear}. Our study indicates that { the} higher-order interactions induce anomalous transitions to synchrony. Unlike the conventional Kuramoto model, higher-order interactions lead to anomalous phenomena such as multistability of full synchronization, incoherent, and two-cluster states, and transitions to synchrony through slow switching and clustering. Phase diagrams of the dynamical regimes are constructed theoretically and verified by direct numerical simulations. We also show that similar transition scenarios are observed even if a small heterogeneity in the oscillators' frequency is included.

\end{abstract}

  \maketitle

\begin{quotation}
Synchronization is a ubiquitous emergent phenomenon in which many coupled units behave in unison. Given the pervasiveness of synchronization, understanding how it is achieved is a fundamental question. In particular, the nature of the interactions among oscillators has strong consequences on the transition to synchronization. To tackle this issue, it is convenient to consider phase models in which each oscillator is described solely in terms of a phase variable. According to phase reduction theory, the phase model captures the dynamics completely when the coupling among the oscillators is sufficiently weak. If one considers only pairwise interactions, the synchronization transition is described by the Kuramoto-type model. In recent years, however, it has been noted that higher-order (many-body) interactions are crucial to fully capture real-world systems. In this paper, we seek to improve the understanding of the impact of higher-order interactions on the synchronization transition. With such a goal, we consider an ensemble of globally coupled identical phase oscillators subject to two-body and three-body interactions{ , including phase lags in both interactions,} derived through phase reduction. We show that the higher-order interactions induce anomalous coexistence of distinct dynamical regimes and transitions to synchrony even in the presence of small heterogeneity. Given that the phase model is derived from phase reduction, its dynamics could be observed in a wide variety of ensembles of coupled nonlinear oscillators.

\end{quotation}
  
\section{Introduction} 

In nature, we recurrently observe the emergence of collective behaviors in systems of interacting dynamical units. One remarkable example of such self-organization is the synchronization of coupled oscillators, observed in circadian rhythms, neuronal dynamics, Josephson junctions, or electric grids, to name a few \cite{PRK01,Win80,Kur84}. Hence, the prediction, control, and understanding of the dynamics of coupled oscillators is a fundamental problem in multiple research fields. 

In order to understand collective synchronization, a common approach is to consider simple phase-oscillator models, where each oscillator is solely described by one degree of freedom, i.e., the phase. The dynamics of the phase models are equivalent to more general systems of nonlinear oscillators, given that the coupling among the oscillators is sufficiently weak, as stated by phase reduction theory \cite{Kur84,nakao16}. 

Despite the power and ductility of such an approach, the classical theory of synchronization is solely based on pairwise interactions, while, in many natural systems, the interactions are intrinsically higher-order (many-body) rather than pairwise \cite{battiston20,bick2023higher}. From ecology \cite{grilli_allesina} to neuroscience \cite{petri2014homological,sizemore2018cliques}, many examples show that a pairwise description is not sufficient to match the theory with observations and, additionally, higher-order interactions appear naturally when phase reduction is performed up to higher orders \cite{leon19,gengel21,leon22a}. 
 Moreover, theoretical studies on consensus \cite{neuhauser2020multibody}, random walks \cite{carletti2020random}, synchronization \cite{gambuzza2021stability,lucas20}, Turing pattern formation \cite{muologallo}, and social contagion \cite{iacopini2019simplicial}, to name a few, showed that higher-order interactions can dramatically affect the global behavior of the system. In particular, it was shown that extensions of the Kuramoto model including higher-order interactions exhibit an explosive transition to synchrony or collective chaos \cite{bick_explosive,millan20,SA19,SA20,leon19,leon22a}.

The goal of this work is to analyze the collective dynamics of the simplest minimal extension of the Kuramoto-type phase model for identical globally coupled oscillators subject to two- and three-body interactions with permutation symmetry{ , including the effect of phase lags}. The simplicity of the model allows us to perform a complete analysis of the phase diagram,
evidencing that higher-order interactions induce anomalous transitions to synchrony. {  We consider those transitions to be anomalous because,} due to the three-body coupling, synchronization is not achieved as in the conventional two-body Kuramoto model or any of its extensions. Instead, we observe either multistability of three states, namely, full synchronization, incoherence, and two-cluster states, or a route to synchronization involving a slow-switching between two clusters. Moreover, we confirm that most of these behaviors, induced by three-body interactions, are robust with respect to the heterogeneity in the oscillators' natural frequencies. Because the model is derived from phase reduction, we expect similar scenarios to be exhibited by a wide variety of systems.

\section{Phase model} 

In the present work, we consider an ensemble of $N$ identical phase oscillators globally coupled through two- and three-body interactions with permutation symmetry:
	\begin{equation}\label{eq.phmodelsum}
	\dot{\theta}_j=\omega+\frac{K_1}{N}\sum_{k=1}^{N}\sin(\theta_k-\theta_j+\alpha)+\frac{K_2}{N^2}\sum_{k,l=1}^{N}\sin(\theta_k+\theta_l-2\theta_j+\beta)
	\end{equation}
for $j=1,...,N$, where $K_1$ and $K_2$ measure the strength of the two- and three-body interactions, while $\alpha$ and $\beta$ are phase lags of the interactions, respectively. Given the symmetry of the model, it is enough to consider only the case with $K_1>0$.

The phase model~\eqref{eq.phmodelsum} is a straightforward extension of the conventional two-body Kuramoto-type model to include three-body interactions that preserve permutation symmetry. This model can be exactly derived by performing phase reduction on the ensemble of Stuart-Landau oscillators with two- and three-body interactions\cite{ashwin16}. { This} derivation can be found in Appendix~\ref{sec.phred}. Moreover, it can also be obtained through phase reduction of ensembles of general limit-cycle oscillators if additional harmonics in the phase-coupling functions are neglected.

Phase models similar to \eqref{eq.phmodelsum} have been extensively studied in the literature, so we briefly discuss them in what follows in order to highlight our findings. If the three-body interactions are removed, i.e., $K_2=0$, we obtain the paradigmatic Kuramoto-Sakaguchi model displaying a transition to synchrony \cite{SK86,Kur84}. On the other hand, if only the three-body interactions are considered, i.e., $K_1=0$, the phase model studied in \cite{KP15} is recovered. In this case, the system exhibits bistability between incoherent and two-cluster states, where the latter represents a state in which each oscillator takes either of two possible phases. The model described by Eq.~\eqref{eq.phmodelsum} with an additional three-body interaction was studied and derived through second-order phase reduction in \cite{leon19,leon22a}, although, in that case, the strength of higher-order interactions was considered to be much smaller than that of the pairwise interactions. Finally, synchronization of model~\eqref{eq.phmodelsum} has been recently studied in the presence of noise \cite{MK23}, bimodally distributed frequencies \cite{car23}, or with inertia \cite{Jaros23}. We remark studies on the three-body interactions with a phase lag in  Eq.~\eqref{eq.phmodelsum} are scarce since the latter is not analyzable in the well-established framework of Watanabe-Strogatz \cite{WS94,WS93} or Ott-Antonsen theory \cite{OA08,OA09}.

Equation~\eqref{eq.phmodelsum} takes a simpler form if, without loss of generality, we fix the frequency to zero, $\omega=0$, by choosing an appropriate rotating frame of reference, and rescaling time as $t\rightarrow K_1 t$. Additionally, we define the Kuramoto order parameter $Re^{i\psi}=\sum_k e^{i\theta_k}/N$ through which the oscillators interact. This allows us to rewrite Eq.~\eqref{eq.phmodelsum} as: 
	\begin{equation}\label{eq.phasemodel}
	\dot{\theta}_j=R\sin(\psi-\theta_j+\alpha)+K R^2\sin(2\psi-2\theta_j+\beta),
	\end{equation}
where $K=K_2/K_1$ measures the ratio between the three- and two-body interactions. In what follows, we will see that three-body (higher-order) interactions give rise to anomalous transitions to synchrony, multistability, and other dynamics that are absent in the original Kuramoto model or its extensions.

\section{Anomalous transitions to synchrony} 

In this section, we perform numerical simulations of the rescaled model \eqref{eq.phasemodel}, evidencing the presence of anomalous transitions to synchrony. Nevertheless, before analyzing the model, let us recall the results for pure pairwise interactions, i.e. $K=0$. In this case, Eq.~\eqref{eq.phasemodel} reduces to the Kuramoto-Sakaguchi model of identical oscillators. It is known that the global attractors of the identical Kuramoto-Sakaguchi model are either the incoherent state, where oscillators are distributed yielding $R=0$, or full synchronization, where oscillators form a point cluster achieving $R=1$. The abrupt transition between those states occurs at $\alpha=\pm\pi/2$.

\begin{figure}
\includegraphics[width=\linewidth]{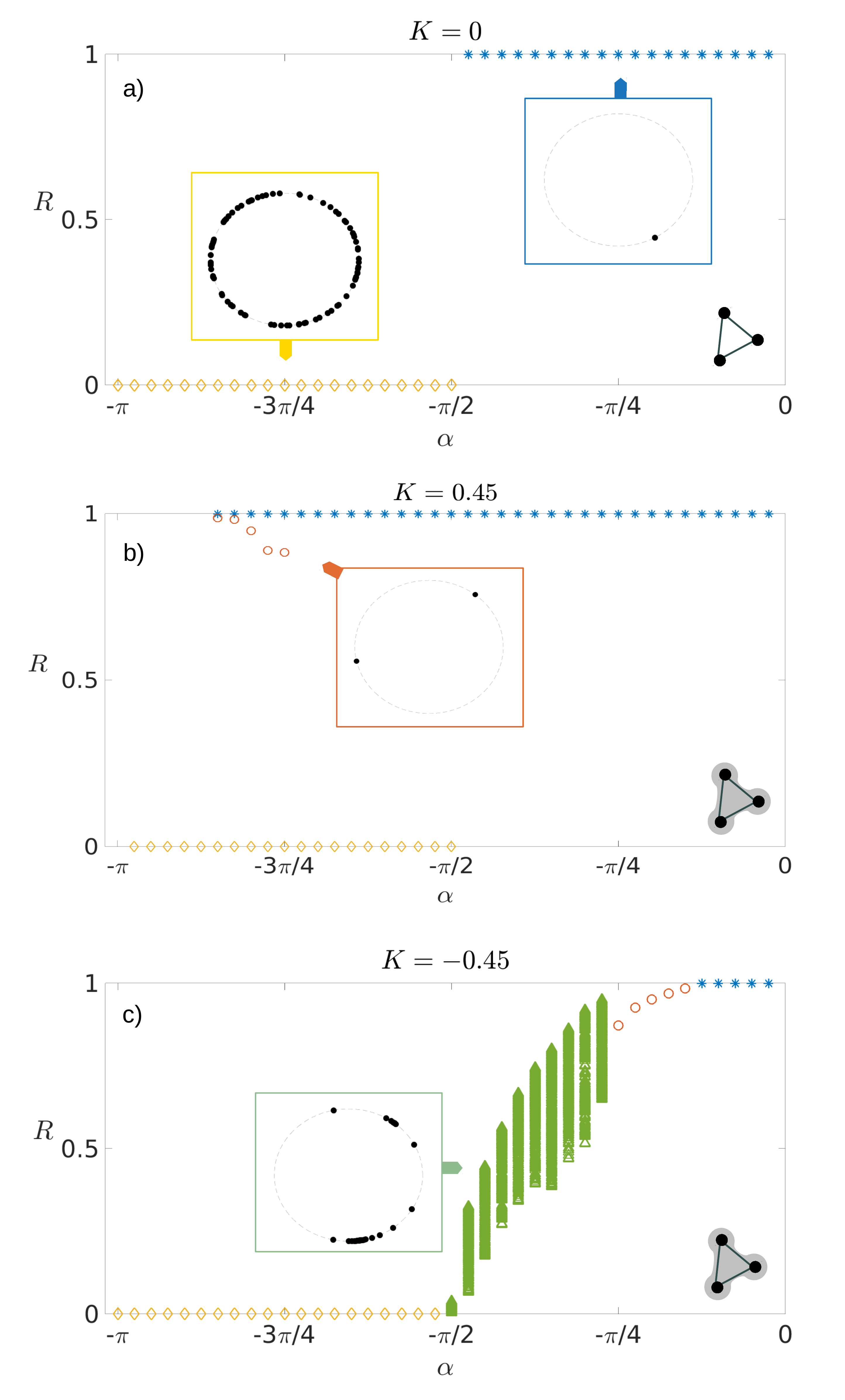}
	\caption
	{Synchronization transitions with pure two-body coupling (a) and with three-body (higher-order) coupling (b,c). Kuramoto order parameter vs. phase lag $\alpha$ for $N=1000$, $\beta=0$, and $K=0$ (a), $K=0.45$ (b), and $K=-0.45$ (c). The yellow (diamonds), blue (stars), red (circles), and green (triangles) indicate the incoherent state, full synchronization, two-cluster state, and slow switching, respectively. A typical snapshot of  each state is shown in the inset to ease understanding of the dynamics.}
	\label{fig.transition}
\end{figure}

In Fig.~\ref{fig.transition} (a), we present the bifurcation diagram of the Kuramoto-Sakaguchi model, where the value of the Kuramoto order parameter $R$, obtained from numerical simulations, is depicted for multiple values of $\alpha$. For visual clarity, yellow diamonds and blue stars are used to indicate the incoherent state and full synchronization, respectively. Additionally, we include snapshots of the oscillator states. As predicted by the theory, the incoherent state becomes unstable at $\alpha=-\pi/2$, giving rise to full synchronization. We note that this transition is different from the smooth second-order transition of the classical Kuramoto model with inhomogeneous frequencies because in our case all oscillators are identical. 

Let us now add the three-body interactions, i.e., $K\not=0$. Note that, since $K_1>0$, the sign of $K$ depends only on the three-body interactions. In Fig.~\ref{fig.transition} (b,c), we plot the order parameter in the steady state after the initial transient versus $\alpha$ for fixed $N=1000$ and $\beta=0$, while choosing the ratio $K=0.45$ in (b) and $K=-0.45$ in (c). These two cases are representative of the dynamics for positive and negative $K$.

First, we focus on the dynamics for $K=0.45$, presented in Fig.~\ref{fig.transition} (b). For any $\alpha<-\pi/2$, we observe that the incoherent state is stable, as in the case without three-body interactions. Nevertheless, the effect of the three-body interactions is remarkable for $\alpha\in(-2.67,-2.35)$, where they induce multistability of the incoherent state, full synchronization, and two-cluster state (red circles). In other words, for the same parameter values, any of these states can be achieved depending on the initial conditions. {  Although multiple two-cluster states are stable for the same $\alpha$, only one of such states is plotted for each $\alpha$ in Fig.~\ref{fig.transition}.} When $\alpha\simeq-2.35$, the numerical simulations indicate that, the two-cluster state loses its stability and the system exhibits bistability between full synchronization and incoherent state. Finally, for $\alpha>-\pi/2$, full synchronization is the only attractor of the system. For larger values of $K$, it is also possible to find regions with bistability between full synchronization and two-cluster state {  and even multistability between the latter states and incoherence} ({ see Fig.~\ref{fig.K12} in appendix~\ref{app.phen}}). The multistability of the system throughout the transition implies that different hysteresis can be detected depending on how the parameters are changed.

The dynamics for $K=-0.45$ are completely different, as we show in Fig.~\ref{fig.transition}(c). For $\alpha<-\pi/2$, the dynamics are similar to the Kuramoto-Sakaguchi model, since the incoherent state is the only attractor. Nonetheless, the system does not achieve full synchronization for $\alpha>-\pi/2$; instead, the order parameter $R$ displays oscillations whose period increases with time ({ see Fig.~\ref{fig.QPSslow} in appendix~\ref{app.phen}}). { To represent these oscillations, in Fig.~\ref{fig.transition}, we plotted a triangle for the value of $R$ observed every 10 time units (t.u.) during 1000 t.u.} This dynamical state is known as slow-switching \cite{hmm93,KK01}, where the system approaches a heteroclinic cycle formed by saddle two-cluster states. The snapshot captures the switching between those saddle two-cluster states. If $\alpha$ is further increased, one of the two-cluster states becomes the only attractor of the system. Finally, we observe that full synchronization is achieved when $\alpha\simeq-0.4$.

Let us remark that the anomalous transition to synchrony, multistability, and slow-switching detected in Fig.~\ref{fig.transition} (b,c) are all caused by the three-body interactions. Although previous studies showed that higher-order interactions were responsible for a wider variety of dynamics { and multistability}~\cite{bick_explosive,millan20,SA19,SA20,leon19,leon22a,lucas20,tanaka11}, the anomalous transitions presented in Fig.~\ref{fig.transition} (b,c) { and the multistability of synchronization, incoherence and two-cluster states} were not reported.

\section{Analytical stability analysis} 

In order to explain the numerical results, we perform an analytical stability analysis of the dynamical regimes displayed by model \eqref{eq.phasemodel}. For mathematical convenience, we consider the thermodynamic limit, $N\rightarrow\infty$, although, with slight modifications, the same analysis could be performed for finite $N$.
 
\begin{figure}
	\includegraphics[width=\linewidth]{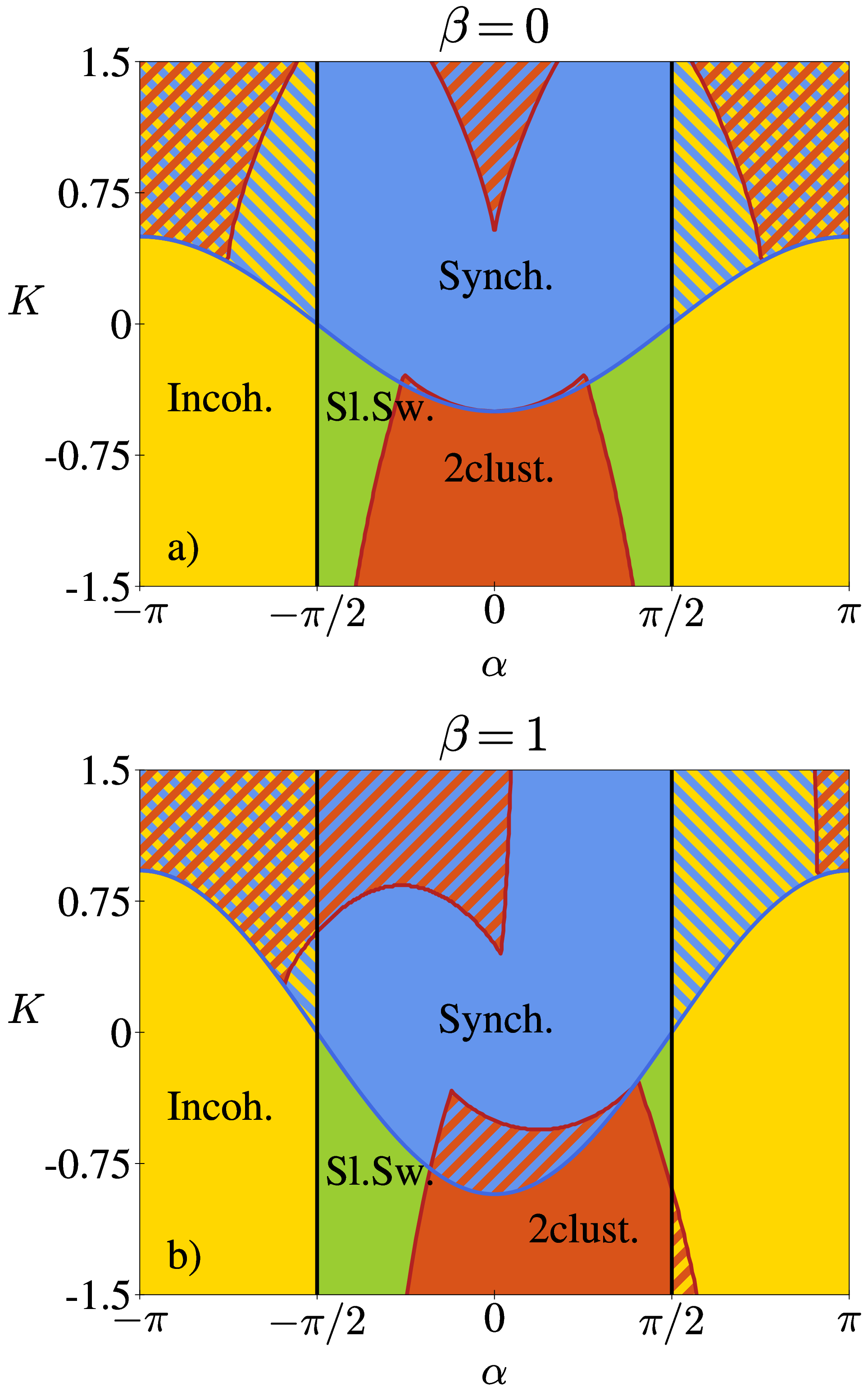}
	\caption
	{Phase diagrams of Eq.~\eqref{eq.phasemodel} for $\beta=0$ (a) and $\beta=1$ (b). In the blue, yellow, red, and green regions, full synchronization, incoherent state, two-cluster state, and slow switching are stable, respectively. The hatching indicates that more than one state is stable using the same color code.}
	\label{fig.phasediag}
\end{figure}

First of all, we study the stability of full synchronization. This state is characterized by all oscillators being located in a point cluster that rotates with frequency $\Omega=\sin\alpha+K\sin\beta$ and thus $R=1$. Although full synchronization is always a solution, linear stability analysis indicates that it becomes stable when
	\begin{equation}
		\cos\alpha+2K\cos\beta>0,
	\end{equation}
depicted in Fig.~\ref{fig.phasediag} (a,b) with a blue line.

Next, we analyze the stability of the incoherent state in which $R=0$.
The linear stability of the incoherent state is the same as in the Kuramoto-Sakaguchi model, because the three-body interaction, proportional to the term $R^2$, vanishes when linearized around $R=0$. Thus, the incoherent state becomes unstable at
	\begin{equation}\label{eq.stabincoh}
		\cos\alpha<0,
	\end{equation}  
depicted by black lines in Fig. \ref{fig.phasediag} (a,b). The independence of Eq.~\eqref{eq.stabincoh} on $K$ explains why the incoherent state is always stable for $\alpha<-\pi/2$ in the numerical simulations. Additionally, we remark that the bistability between the incoherent state and full synchronization is caused by the fact that the stability of full synchronization depends on $K$ while the stability of the incoherent state does not.

When the incoherent state becomes unstable, quasi-periodic partial synchrony (QPS) generically appears \cite{PR15}. QPS is a state in which the Kuramoto order parameter rotates uniformly while individual oscillators behave quasi-periodically. This state is not shown in Fig.~\ref{fig.transition} because it is an unstable saddle for the present model. However, because the saddle QPS is weakly unstable, during the initial transient, the system spends a large amount of time close to this state ({ see Fig.~\ref{fig.QPSslow} in appendix~\ref{app.phen}}).

Finally, we analyze the stability of the two-cluster state following Ref.~\cite{hmm93,KK01,leon19}. First, we note that the two-cluster state is not a single configuration of oscillators but a family of configurations; each two-cluster state is determined by the fraction $p$ of the oscillators in the first cluster and the distance (phase difference) $\Delta$ between the first and second clusters. We can obtain the evolution equation for $\Delta$ for each value of $p$, whose fixed points correspond to the possible two-cluster states. To obtain the stability of these solutions, it is enough to consider the evolution of three types of variations: the variation of $\Delta$ and the variations of one oscillator in the first or second cluster. If those variations decay, the two-cluster state with the given $\Delta$ and $p$ is stable. The region where the two-cluster state is stable is then determined and depicted in red in Fig.~\ref{fig.phasediag} (a,b). See Appendix \ref{sec.appcluster} for thorough details of the stability analysis. { We remark that, as shown in the inset of Fig.~\ref{fig.transition}, generally $\Delta\not=\pi$, however, in the restricted and more symmetric case of positive $K$ and $\alpha=\beta=0$, the only stable two-cluster state has $\Delta=\pi$, similarly to \cite{KP15,SA19,SOR11,OKU93,tanaka11}}.

{ Even} when the two-cluster state is unstable, it is possible to find another dynamical state: slow switching \cite{HM01,KK01}. In this state, two saddle two-cluster states form a heteroclinic cycle and the system approaches this cycle while switching between both unstable two-cluster states. The conditions for such slow switching to be stable were derived in \cite{KK01}, see also Appendix \ref{sec.appslow}. The analysis indicates that slow switching can only be realized when all other states are unstable.

In Fig.~\ref{fig.phasediag} (a), we depict the phase diagram for $\beta=0$. In the blue, yellow, red, and green regions, full synchronization, incoherent state, two-cluster state, and slow switching are stable, respectively. The hatched regions indicate multistability following the same color code. This figure evidences that the system displays multi-stability for wide parameter ranges. Additionally, we remark that, given the intricate phase diagram, different paths along the parameter space will give rise to different anomalous transitions to synchrony. We emphasize that the present phase diagram has been computed analytically, explaining the anomalous transitions observed in Fig.~\ref{fig.transition}. The phase diagram obtained by direct numerical simulations is in excellent agreement with Fig.~\ref{fig.phasediag}~(a) as reported in Appendix \ref{sec.appphase}.

For the sake of completeness, we have also analyzed the model for other values of $\beta$. Although some quantitative changes are observed, the dynamics and bifurcations are qualitatively the same. As a particular example, in Fig.~\ref{fig.phasediag} (b), we depict the phase diagram for $\beta=1$. By comparing it with the case with $\beta=0$, we observe that the stability boundaries are shifted and the stability regions for the two-cluster state are deformed, but no new dynamical regimes or transitions arise. We highlight here that, although the dynamics are equivalent, the value of $\beta$ might be important in applications. For example, the wide region of bistability between full synchronization and the two-cluster state for negative $K$ in Fig.~\ref{fig.phasediag} (b) is almost inappreciable for $\beta=0$ in Fig.~\ref{fig.phasediag} (a).

\section{Effects of heterogeneity} \label{sec.heter}

We have so far assumed all the oscillators to be identical. Although a complete analysis of the effect of heterogeneity is beyond the scope of this work, one may ask whether the above results are robust against small heterogeneity. In order to answer this question, we consider the natural frequency of each oscillator to be drawn from a normal distribution with zero mean and standard deviation $\sigma=0.05$, $\mathcal{N}(0,0.05)$, and perform a numerical study analogous to the above one.

The first consequence of the heterogeneity is that full synchronization is no longer possible since the heterogeneity prevents all oscillators from forming a point cluster. Nevertheless, the system evolves to partial synchrony, where some oscillators are synchronized while others are drifting.

For positive values of $K$, the addition of small heterogeneity produces some quantitative differences in the stability boundaries; however, it is still possible to find the regions with multistability of partial synchrony, incoherent state, and two-cluster state. This means that higher-order interactions promote anomalous transitions to synchrony and give rise to wide regions of multistability even in the presence of small heterogeneity.

The effect of heterogeneity when $K<0$ is more noticeable. In fact, in the regions where we observed slow switching in the identical case, the system now displays partial synchrony. This means that the heterogeneity induces partial synchrony due to the three-body interactions. This phenomenon can be understood as the heterogeneity stabilizing the saddle QPS, similar to the stabilization of QPS by noise observed in \cite{CP17}.

\section{Conclusions} 
In this work, we have studied the simplest general model of globally coupled identical oscillators subject to pairwise and higher-order interactions with permutation symmetry.
The considered model is a natural extension of the Kuramoto{ -Sakaguchi} model with an additional three-body interaction {  with a phase lag} that can be derived from an ensemble of Stuart-Landau oscillators with higher-order interactions. As we have shown, the three-body coupling plays a crucial role in the dynamics of the system{ ; when the strength of the couplings or the phase lags are varied, the higher-order interactions} give rise to anomalous transitions to synchrony and promote the multistability of synchronous, incoherent, and two-cluster states. These results have been obtained numerically and corroborated analytically through a stability analysis.

We stress that the anomalous transitions to synchrony and multistability are not degenerate scenarios caused by the fact that all oscillators are identical. In fact, we numerically observed the system displaying similar behaviors even when small heterogeneity is included. Thus, given that the phase model we analyzed is derived from a general system of coupled oscillators through phase reduction, we believe that the complex dynamical scenarios described in this study can be achieved in a wide variety of systems.

The results included in this work are a fundamental step forward in our understanding of the effect of higher-order interactions, which find applications in many research fields. We believe that our analysis paves the way for future studies involving the presence of noise, heterogeneity, or different higher-order network topologies \cite{skardal23,malizia23,Zhang23,gallo2022synchronization}. \\

\section*{Acknowledgements} 
I.L. and H.N. acknowledge JSPS KAKENHI JP22K11919, JP22H00516, and JST CREST JP-MJCR1913 for financial support. During the realization of this work, R.M. was supported by an FRIA-FNRS Fellowship, funded by the Walloon Region, Grant FC 33443. R.M. also acknowledges the Erasmus+ and the Mobility Out program of the FNRS for funding his visit in the group of H.N.

 \section*{Additional Information}
The address of I.L. when the review was performed was \textit{Department of Applied Mathematics and Computer Science, Universidad de Cantabria, Santander, Spain}. \\
The current address of R.M. is \textit{Department of Systems and Control Engineering, Tokyo Institute of Technology, Tokyo, Japan}.

\section*{DATA AVAILABILITY}
 
 The data that support the findings of this study are available within the article.

\appendix

\section{Phase reduction}\label{sec.phred}

\label{appA:alg_topology}
\setcounter{equation}{0}
\renewcommand{\theequation}{A\arabic{equation}}
This section is devoted to showing how the phase model \eqref{eq.phmodelsum} is obtained by performing phase reduction on an ensemble of Stuart-Landau oscillators. We consider $N$ Stuart-Landau oscillators globally coupled with two- and three-body interactions:

\begin{eqnarray}\label{eq.SL}
\dot{W}_j &=& (1+i\tilde{\omega})W_j - (1 + i c_2) |W_j|^2 W_j\nonumber\\
&+& \frac{\kappa_1(1+ic_1)}{N} \sum_{k=1}^N (W_k - W_j)\\
&+& \frac{\kappa_2(1+ic_3)}{N^2} \sum_{k=1}^N \sum_{\ell=1}^N ( W_k W_\ell W_j^* - |W_j|^2 W_j ),\nonumber
\end{eqnarray}
where $W_j$ is the oscillator's complex variable, $\Tilde{\omega}-c_2$ is the frequency of the oscillator, $c_2$ is the non-isochronicity parameter, $\kappa_1$ and $\kappa_2$ are the strength of the two- and three-body interactions while $c_1$ and $c_3$ are the `reactivities' of the two- and three-body coupling, respectively. 
We note that the three-body interaction in the above model is the simplest case that satisfies the following conditions: (i) symmetric with respect to permutations of the interacting oscillators $k$ and $l$, (ii) symmetric with respect to rotation of all oscillators on the complex plane, i.e., $W_j \to W_j e^{i \chi}$ where $\chi$ is a real number, and (iii) vanishing when all oscillators synchronize.

The Stuart-Landau oscillator is the normal form of the Hopf bifurcation, and thus, model Eq.~\eqref{eq.SL} approximately describes the dynamics of a population of oscillators close to a Hopf bifurcation. The model can be obtained by the center-manifold reduction of a general model of nonlinear oscillators~\cite{Kur84}, where all terms and parameters considered naturally appear as the result of the reduction. We remark that model \eqref{eq.SL} contains the simplest higher-order interaction that preserves the permutation symmetry of three oscillators, that is, the interaction is invariant if the subindex $k$ and $l$ are exchanged. Additionally, because non-resonant terms are eliminated in the center-manifold reduction, the model presents rotational symmetry. 

In this appendix, we perform phase reduction following the standard phase-reduction theory~\cite{Kur84,nakao16}, applicable to any oscillator. However, for the specific model \eqref{eq.SL}, we could also follow Ref.~\cite{ashwin16}, where the symmetries are required to obtain the phase-reduced model. In order to perform phase reduction, it is convenient to make a change to Cartesian coordinates, $W_j=x_j+iy_j$, and rewrite Eq. \eqref{eq.SL} in the form:
\begin{eqnarray}
    \dot{\boldsymbol{X}}_j&=&\boldsymbol{F}(\boldsymbol{X}_j)+\frac{\kappa_1}{N}\sum_{k=1}^{N}\boldsymbol{p}_1(\boldsymbol{X}_k,\boldsymbol{X}_j)\nonumber\\
    &+&\frac{\kappa_2}{N^2}\sum_{k,l=1}^{N}\boldsymbol{p}_2(\boldsymbol{X}_k,\boldsymbol{X}_l,\boldsymbol{X}_j),
\end{eqnarray}
where $\boldsymbol{X}_j=(x_j,y_j)$ and $\boldsymbol{p}_1$ and $\boldsymbol{p}_2$ represent two- and three-body interactions, respectively. Phase reduction theory states that the evolution of the phase $\theta_j$ of the oscillator $j$ is represented by the natural frequency plus the product of the phase response function~\cite{Kur84,nakao16} and the interaction functions evaluated on the limit cycle. This gives
    \begin{eqnarray}\label{eq:app_phred}
        \dot{\theta}_j&=&\Tilde{\omega}-c_2+\frac{\kappa_1}{N}\sum_{k=1}^{N}\boldsymbol{Z}(\theta_j)\cdot\boldsymbol{p}_1(\theta_k,\theta_j) \nonumber \\
    & & +\frac{\kappa_2}{N^2}\sum_{k,l=1}^{N}\boldsymbol{Z}(\theta_j)\cdot\boldsymbol{p}_2(\theta_k,\theta_l,\theta_j),
    \end{eqnarray}
where $\boldsymbol{Z}(\theta)=(-\sin\theta-c_2\cos\theta, \cos\theta-c_2\sin\theta)$ is the phase response function of the Stuart-Landau oscillator and we have evaluated the oscillator states $\boldsymbol{X}_{i}$ in $\boldsymbol{p}_{1,2}$ on the limit cycle $\boldsymbol{X}_i=(\cos\theta_i, \sin\theta_i)$ where $i=k, l, j$.

Evaluating Eq. \eqref{eq:app_phred} yields the phase model described by Eq.~\eqref{eq.phmodelsum}, where the constants $\alpha$, $\beta$, $K_1$, $K_2$, and $\omega$ take the values
\begin{eqnarray}
    \alpha&=&\arg[1+c_1 c_2+(c_1-c_2)i],\\
    \beta&=&\arg[1+c_3 c_2+(c_3-c_2)i],\\
    K_1&=&\kappa_1\sqrt{(1+c_1^2)(1+c_2^2)},\\
    K_2&=&\kappa_2\sqrt{(1+c_3^2)(1+c_2^2)},\\
    \omega&=&\tilde{\omega}-c_2-\kappa_1(c_1-c_2)-\kappa_2(c_3-c_2).
\end{eqnarray}
We note that averaging is not necessary since \eqref{eq:app_phred} yields only resonant terms due to the rotational symmetry of \eqref{eq.SL}.

\section{Other phenomenology}\label{app.phen}
\setcounter{figure}{0}
\renewcommand{\thefigure}{B\arabic{figure}}

We devote this section to showing in more detail some of the observed phenomenology.

\begin{figure}[h]
\includegraphics[width=\linewidth]{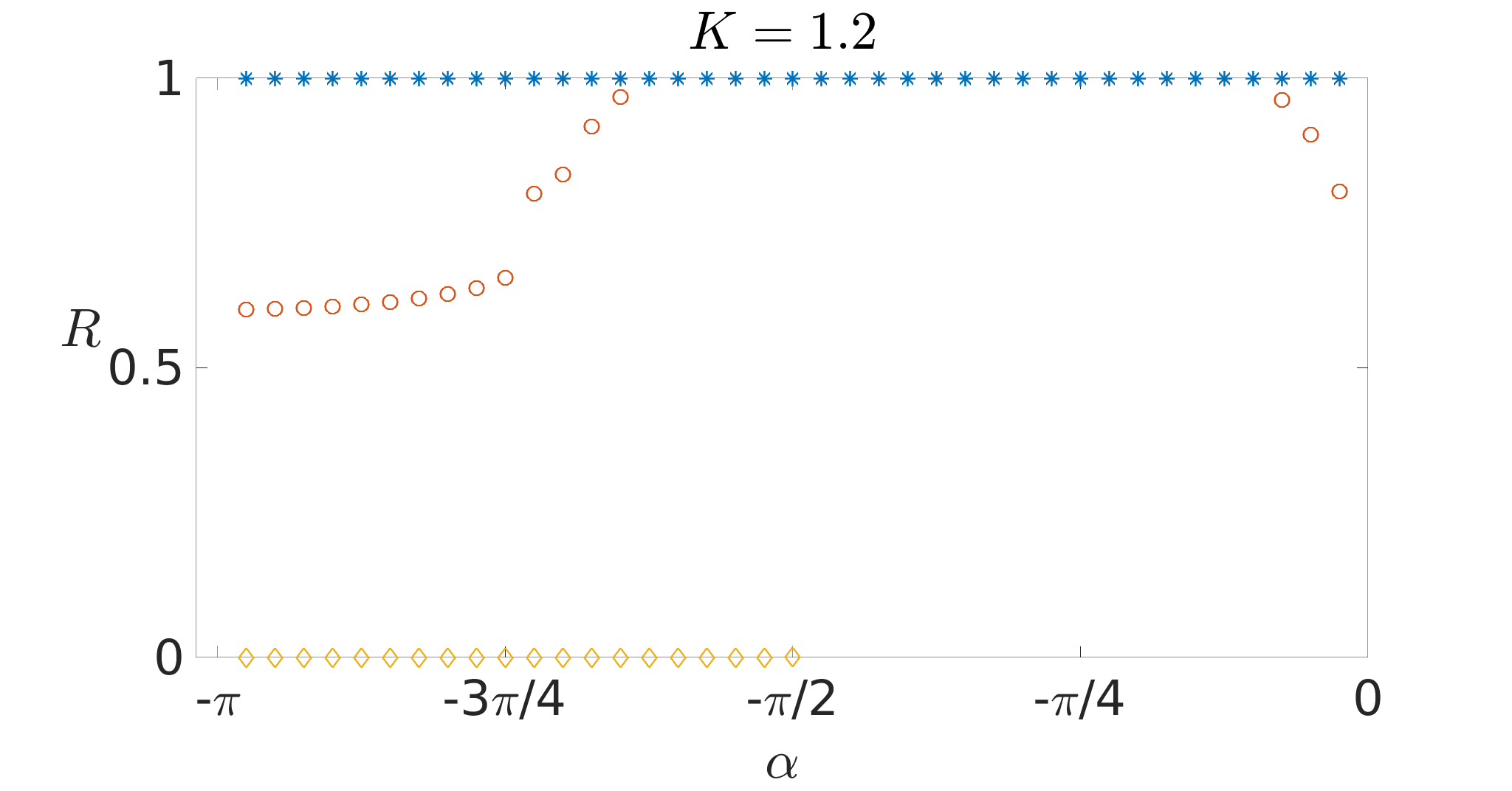}
\caption{Kuramoto order parameter vs. phase lag $\alpha$ for $N=1000$, $\beta=0$, and $K=1.2$. The yellow (diamonds), blue (stars) and red (circles) indicate the incoherent state, full synchronization and two-cluster state, respectively.}\label{fig.K12}
\end{figure}

In Fig.~\ref{fig.K12}, we represent the value of the Kuramoto order parameter $R$ versus $\alpha$ for $K=1.2$ and $\alpha=0$, similarly to what was done in Fig~\ref{fig.transition}. For this larger value of $K$, we can observe bistability between full synchronization and two-cluster states.

\begin{figure}[h]
\includegraphics[width=\linewidth]{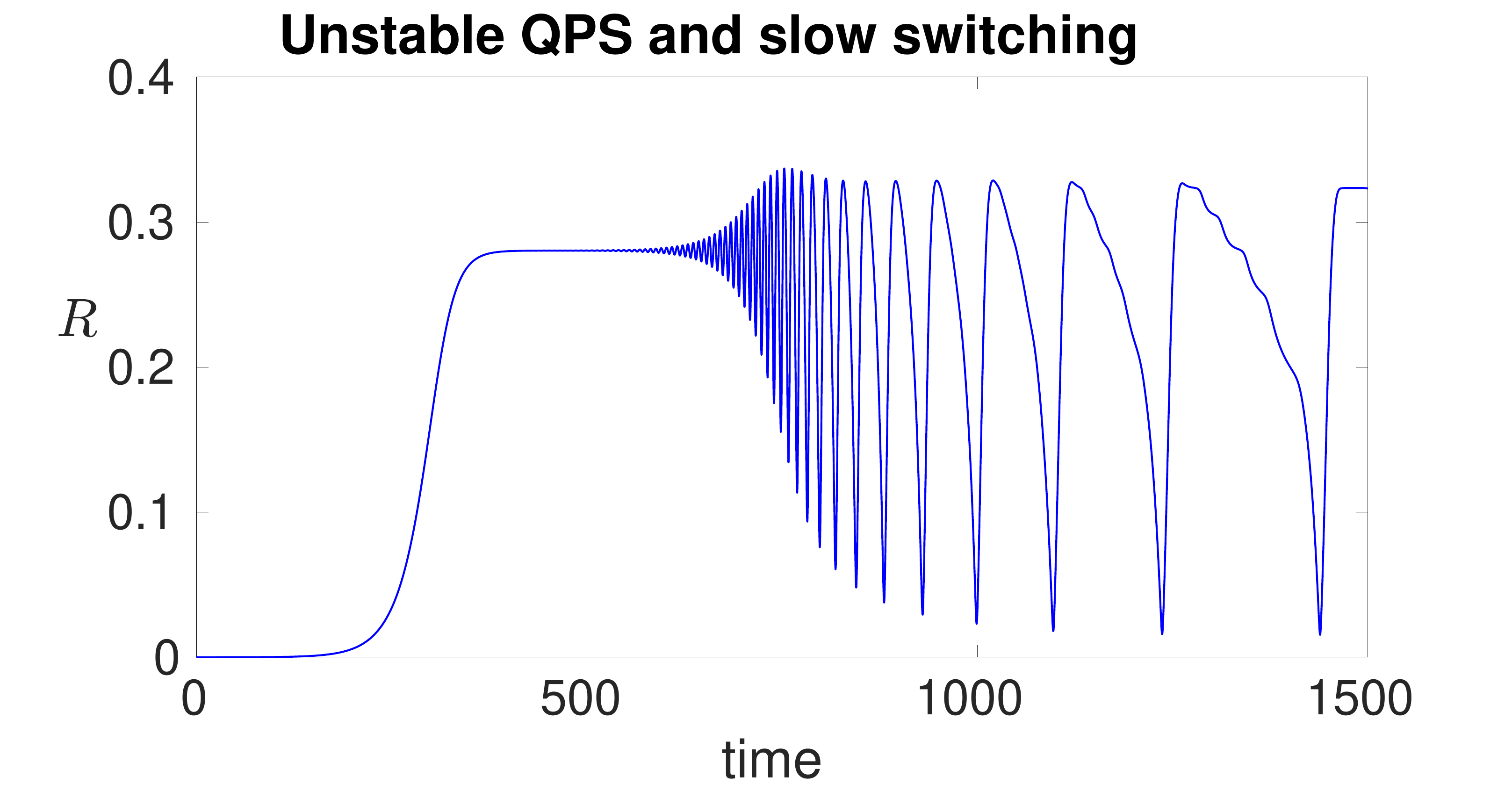}
\caption{Time series of Kuramoto order parameter $R$ of the system Eq.~\eqref{eq.phasemodel} in the main text for $K=-0.45$ , $\alpha=-1.5$, and $\beta=0$ initialized close to the incoherent state. We observe the system get close to an unstable QPS where the order parameter rotates uniformly (hence the time series shows a plateau). The system displays slow switching after it departs from the unstable QPS.}\label{fig.QPSslow}
\end{figure}

In Fig~\ref{fig.QPSslow}, we illustrate the dynamics of the system for $K=-0.45$, $\alpha=-1.5$, and $\beta=0$, where slow switching is the only stable state. In the figure, the order parameter initially grows and stays close to the saddle QPS (plateau) for a long time and then it approaches the heteroclinic limit cycle of the slow switching. We can observe how the period of the oscillations grows.
\section{Stability of two-cluster state}\label{sec.appcluster}
\setcounter{equation}{0}
\renewcommand{\theequation}{B\arabic{equation}}

We devote this section to performing stability analysis of the two-cluster state displayed by the phase model~\eqref{eq.phasemodel}, following Refs.~\cite{hmm93,KK01,leon19}. First of all, we rewrite the model in the compact form:
\begin{equation}
	\dot{\theta}_j=\frac{1}{N}\sum_{k=1}^{N}\Gamma(\theta_k-\theta_j)+ \frac{1}{N^2}\sum_{k,m=1}^{N}g_2(\theta_k+\theta_m-2\theta_j),
\end{equation}
where $\Gamma(x)=\sin(x+\alpha)$ and $g_2(x)=K \sin(x+\beta)$.

We can characterize the two-cluster state by a fraction of oscillators $p$, forming cluster $A$ at the phase $\theta_A$, and the other fraction $(p-1)$, forming cluster $B$ at the phase $\theta_B$. The evolution of the phases $\theta_A$ and $\theta_B$ of the clusters obeys
\begin{eqnarray}
	\dot{\theta}_A&=&p\Gamma(0)+(1-p)\Gamma(\theta_B-\theta_A)+p^2g_2(0)\\
         &+&2p(1-p)g_2(\theta_B-\theta_A)+(1-p)^2g_2(2\theta_B-2\theta_A) ,\nonumber\\
	\dot{\theta}_B&=&(1-p)\Gamma(0)+p\Gamma(\theta_A-\theta_B)+(1-p)^2g_2(0)\\
        &+&2p(1-p)g_2(\theta_A-\theta_B)+p^2g_2(2\theta_A-2\theta_B).\nonumber
\end{eqnarray}

If the two-cluster state is stable, the distance between the clusters, $\Delta=\theta_A-\theta_B$, is constant. The evolution of the distance $\Delta$ is given by
\begin{multline}\label{eq.delta}
	\dot{\Delta}=(2p-1)\Gamma(0)+(1-p)\Gamma(-\Delta)-p\Gamma(\Delta)\\
 +(2p-1)g_2(0)+2p(1-p)[g_2(-\Delta)-g_2(\Delta)]\\
 +(1-p)^2g_2(-2\Delta)-p^2g_2(2\Delta).
\end{multline}
The pairs of $(p,\Delta)$, with $p\in(0,1)$ and $\Delta\in[0,2\pi)$, such that the right hand side of \eqref{eq.delta} is zero are the possible two-cluster states. However, this condition only implies the existence of the two-cluster states, not their stability. 
The stability of the two-cluster state can be computed by decomposing small variations from the two-cluster state into three orthogonal modes~\cite{GHS+92,KK01}; one mode corresponds to the phase locking of the two clusters while the other two modes capture the disintegration of clusters $A$ and $B$, respectively. The decay of these modes is characterized by eigenvalues $\lambda_L$, $\lambda_A$, and $\lambda_B$, respectively. 

We first consider the stability of the phase locking of the two clusters by studying the variation of their distance (phase difference) $\Delta$. Linearizing \eqref{eq.delta} around the fixed point $(p,\Delta)$, such variation grows with the exponent
\begin{multline}
	\lambda_L=-(1-p)\Gamma'(-\Delta)-p\Gamma'(\Delta)
        -2p(1-p)[g_2'(-\Delta)+g_2'(\Delta)]\\-2(1-p)^2g_2'(-2\Delta)-2p^2g_2'(2\Delta),
\end{multline}
where $'$ indicates the derivative.

To analyze the stability of the cluster $A$ against disintegration, we compute the evolution of the variation of a single oscillator from the cluster $A$. This variation will grow with the exponent
\begin{multline}\label{eq.LA}
	\lambda_A=-p\Gamma'(0)-(1-p)\Gamma'(-\Delta)-2p^2g_2'(0)\\
        -4p(1-p)g_2'(-\Delta)-2(1-p)^2g_2'(-2\Delta) .
\end{multline}
The exponent associated with the disintegration of cluster $B$ is obtained by changing $p\rightarrow(1-p)$ and $\Delta\rightarrow-\Delta$ in Eq.~\eqref{eq.LA}. When all three eigenvalues are negative, the corresponding two-cluster state is stable.

\section{Stability of slow switching}\label{sec.appslow}
\setcounter{equation}{0}
\renewcommand{\theequation}{C\arabic{equation}}

In this section, we explain the explicit conditions for slow switching to be stable. As previously stated, slow switching is a dynamical state in which the system approaches a heteroclinic cycle formed by two unstable two-cluster states. This state is stable if the following conditions hold \cite{KK01}:
	\begin{itemize}
	\item There is at least one value of $p$ such that three different two-cluster state exist. The distances $\Delta_{1,2,3}$ between the clusters in those states are ordered as $0<\Delta_{1}<\Delta_2<\Delta_{3}<2\pi$ and their eigenvalues are denoted as $\lambda_{L,A,B}^{1,2,3}$. 
	\item Full synchronization is unstable.
	\item $\lambda_L^{2}>0$ while $\lambda_L^{1}<0$ and $\lambda_L^{3}<0$.
	\item $\lambda_A^{1}>0$ and $\lambda_B^{1}<0$ while $\lambda_A^{3}<0$ and $\lambda_B^{3}>0$.
\end{itemize}
These conditions ensure the two-cluster states characterized by $\Delta_1$ and $\Delta_3$ form a stable heteroclinic cycle.
For the model studied in the main text, these conditions are only satisfied when full synchronization, incoherence, and two-cluster states are unstable.

\section{Comparison between numerical and analytical phase diagram}\label{sec.appphase}
\setcounter{equation}{0}
\renewcommand{\theequation}{E\arabic{equation}}
\setcounter{figure}{0}
\renewcommand{\thefigure}{E\arabic{figure}}

In this section, we compare the numerically obtained phase diagram of model~\eqref{eq.phasemodel} with the analytical results.

\begin{figure*}[h!]
\includegraphics[width=\textwidth]{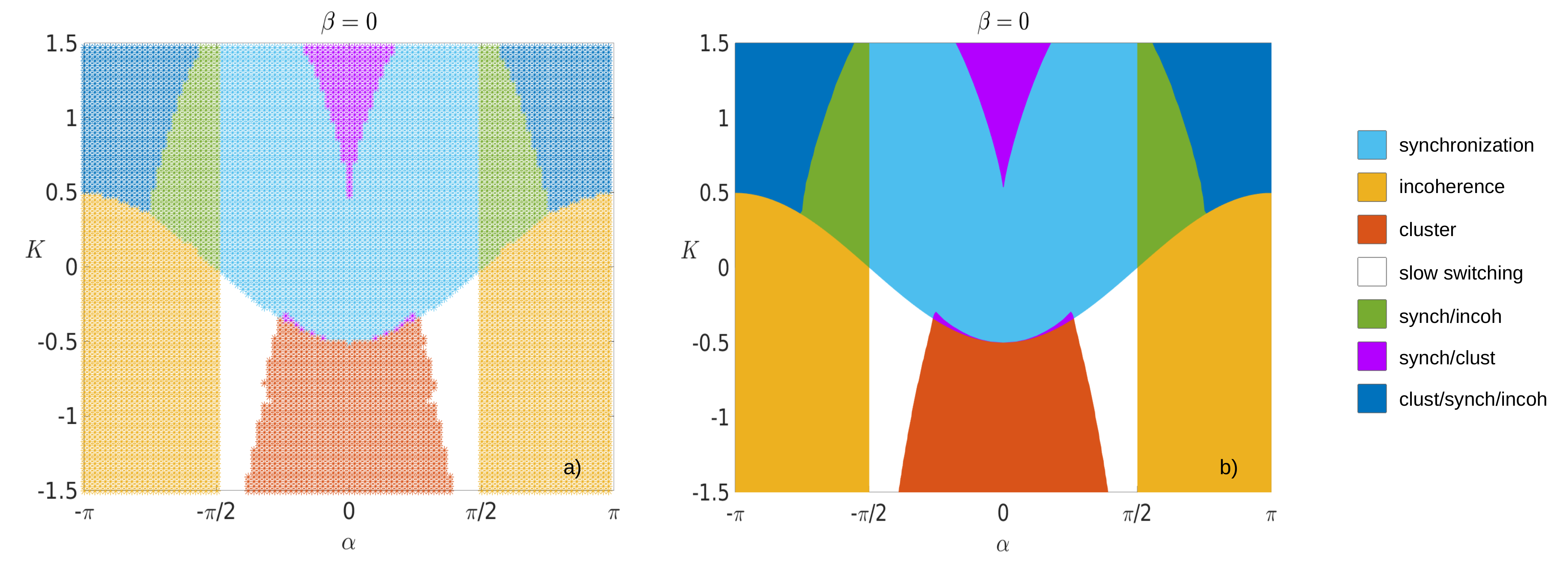}
	\caption
	{Phase diagrams of Eq.~\eqref{eq.phasemodel} for $\beta=0$ obtained by direct numerical simulations (a) and linear stability analysis (b). The light blue, yellow, red, or white color indicates the incoherent state, full synchronization, two-cluster state, or slow switching is stable, respectively, while violet, green, or dark blue indicates bistability of full synchronization and the two-cluster state, bistability of incoherence and full synchronization, and multistability of incoherence, full synchronization, and the two-cluster state, respectively.}
	\label{fig.phasediagapp}
\end{figure*}

In Fig.~\ref{fig.phasediagapp} (a), we depict the phase diagram obtained by direct numerical simulations with $N=1000$ oscillators and $\beta=0$. We initialized the system close to each of the possible states and plot a blue, yellow, red, or white point if the incoherent state, full synchronization, two-cluster state, or slow switching was stable, respectively. If two or more states were stable, a violet, green, or dark blue point was respectively used to denote bistability of full synchronization and the two-cluster state, bistability of incoherence and full synchronization, and multistability of incoherence, synchrony, and the two-cluster states.

In Fig.~\ref{fig.phasediagapp} (b), we have replotted the analytically obtained phase diagram of Fig.~\ref{fig.phasediag} (a) with the color code of Fig.~\ref{fig.phasediagapp} (a). The comparison of Fig.~\ref{fig.phasediagapp} (a) and (b) gives the evidence of the excellent agreement between the numerical and analytical results.

{ The numerical simulations were performed using a fourth order Runge-Kutta algorithm with a constant time step of $dt=0.01$ for $N=1000$ oscillators for a total time of $t=10^4$ t.u.  Numerically, slow-switching stops when the system is too close to a saddle point, due to finite precision. We used the presence of big oscillations in the order parameter previous to this stop to determine the existence of slow-switching.}

\end{document}